\documentclass[preprint2]{aastex}

\shortauthors{Ibungochouba and Ibohal}

\begin{document}

\title{Hawking's radiation  in non-stationary rotating de Sitter background}



\author{Ng. Ibohal and T. Ibungochouba\altaffilmark{1}}
\affil{Department of Mathematics\\
Manipur University, Canchipur, Imphal-795003\\
Manipur, India}
\email{ngibohal@iucaa.ernet.in and ibungochouba@rediffmail.com }
\altaffiltext{1}{United College, Lambung, Chandel (Manipur),  PIN-795127,
Manipur, India}

\begin{abstract}
Hawking's radiation effect of Klein-Gordon scalar field, Dirac particles and Maxwell's electromagnetic field in the non-stationary rotating de Sitter cosmological space-time is investigated by using a method of  generalized tortoise co-ordinates transformation. The locations and the temperatures of the cosmological horizons of the non-stationary rotating de Sitter model are derived. It is found that the locations and the temperatures of the rotating cosmological model depend not only on the time but also on the angle. The stress-energy regularization techniques are applied to the two dimensional analog of the de Sitter  metrics and the calculated stress-energy tensor contains the thermal radiation effect.
\end{abstract}
\keywords{Rotating non-stationary de Sitter cosmological model;  Hawking radiation; generalized tortoise coordinate transformation}

\section{Introduction}

The black hole thermal radiation technique of quantum field theory was initially discovered by \citet{haw74,ha75}. The quantum thermal radiation due to black hole has been investigated by different authors in different types of space-time, such as  Kerr-Newman \citep{dam76}, Kerr-Newman-Kasuya \citep{ahmmon95}, Kerr \citep{zhao83}, Vaidya-Schwarzchild-de Sitter \citep{da93}  and NUT-Kerr-Newman-de Sitter \citep{ahm91} space-time. \citet{zha92} proposed a suitable method known as generalized tortoise co-ordinate transformation (GTCT) which can determine the exact value of both the location and the surface gravity of the horizons of non-stationary black hole. The idea of GTCT is to reduce the scalar field equations, like Klein-Gordon or Dirac, in a black hole background to a standard wave equation near the horizon. \citet{lizh93} investigated the Hawking's effect of a Dirac particles in a non-static case. The study of Hawking evaporation of Dirac particles in a non-stationary axisymmetric black hole is complicated because there is no unique method for the separation of variables for the radial and angular parts involved in the Chandrasekhar-Dirac equations \citep{chan83} of non-stationary axisymmetric space-time. \citet{wocai01} studied Hawking's effect on various non-stationary space-times, rotating Vaidya, non-rotating Vaidya de Sitter (2002a) and rotating Vaidya-Bonnor (2002b). For an evaporating black hole, the vacuum expectation value of the renormalized energy momentum tensor determines the location and temperature of the event horizon \citep{his81}. It is well known that the de Sitter line element is a maximally symmetric curve space-time. The non-rotating de Sitter space is conformally flat $C_{abcd}=0$ space-time with constant curvature $R_{abcd}=(\Lambda/3)(g_{ac}g_{bd}-g_{ad}g_{bc})$ \citep{hawell73}.

The aim of this paper is to investigate the Hawking's radiation effect of Klein-Gordon scalar field, Dirac particles and Maxwell's electromagnetic field in non-stationary rotating de Sitter cosmological space-time. The non-stationary rotating de Sitter solution is algebraically special in Petrov classification of space-times \citep{ibo09}, whereas the stationary rotating de Sitter is Petrov type $D$ possessing a geodesic, shear-free, expanding  as well as non-zero twist null vector $\ell_a$, which is one of the repeated principal null directions of the Weyl curvature tensor \citep{ibo05}.
For the sake of completeness of the presentation, we cite that the non-rotating stationary \citep{hawell73} and non-stationary de Sitter  \citep{ibo09} models are conformally flat $(C_{abcd}=0)$ solutions of Einstein's field equations. It is worth to mention that the rotating de Sitter models, stationary as well as non-stationary, possess four roots of the polynomial $\Delta \equiv r^2 - \Lambda(u) r^4/3+a^2=0$, which describes the singularity of the models, whereas most known black holes (rotating and non-rotating) have two roots of the polynomial. It is noted that the non-rotating de Sitter spaces, stationary and non-stationary, with $a=0$ have two roots of the polynomial $\Delta =0$. Gibbons and Hawking (1977) have suggested that the de Sitter cosmological event horizon has formal similarities with a black hole event horizon. Therefore, it would be interesting to investigate the radiation spectra due to the Klein-Gordon scalar field, Dirac particles and Maxwell's electromagnetic field in the rotating de Sitter space having  four roots of the polynomial describing the cosmological horizons.

The outlines of the paper are as follows: In Section 2, we reintroduce the non-stationary rotating de Sitter solution with cosmological function $\Lambda(u)$ in Newman-Penrose (NP) formalism to be used in the study of scalar field equations in the next sections. The NP spin coefficients of de Sitter solution are helpful to study of the second order wave equations as seen in the remaining sections.  In this section we also introduce an improved GTCT transformation in the  non-stationary rotating de Sitter cosmological background having four roots of the polynomial, which describes the singularities of the solutions. The locations of the horizons near $r=r_j, j=1,2,3,4$ satisfying the null surface condition are given for discussion in the next sections. The Klien-Gordon equation of scalar field is analyzed in Section 3. The adjustable parameters $\kappa_j$ appearing in the GTCT near the horizons are found for the  Chandrasekhar-Dirac and Maxwell's equations in Section 4 and 5 respectively in the rotating de Sitter background. In Section 6, we recast each second order equation to a standard wave equation of the Klein-Gordon scalar field, Dirac particles and Maxwell's electromagnetic field, and study the thermal radiation spectra of the fields by expressing the exact values of Hawking temperature. Section 7 deals the calculation of the stress-energy tensor for two dimensional rotating de Sitter metric, and it is found to contain the thermal radiation in the time-time component $g_{uu}$ of the metric tensor. Section 8 deals with the conclusion and remarks of the results obtained in the paper. The presentation of this paper is essentially based on the Newman-Penrose spin-coefficient formalism \citep{newpen62} in $(+,-,-,-)$ signature.

\section{Non-stationary rotating de Sitter solution and generalized tortoise co-ordinate transformation}

Here it is convenient to reintroduce the non-stationary rotating de Sitter \citep{ibo09} in NP formalism for the study of Hawking's radiation effect of Klein-Gordon scalar field, Dirac particles and Maxwell's electromagnetic field.
The line element, describing  non-stationary rotating de Sitter
solution with a cosmological function $\Lambda(u)$ in
the null co-ordinates system $(u, r, \theta, \phi)$ is
\begin{eqnarray}
ds^2&=&\Big[1-\frac{\Lambda(u) r^4}{3R^2}\Big]du^2 +2du\,dr \cr
&&+\frac{2a\Lambda(u) r^4}{3R^2}{\ sin}^2\theta\,du\,d\phi -2a\,{\rm
sin}^2\theta\,dr\,d\phi \cr &&-R^2d\theta^2 -\{(r^2+a^2)^2\cr&&-\Delta
a^2\,{\rm sin}^2\theta\}\,R^{-2}{\rm sin}^2\theta\,d\phi^2,
\end{eqnarray}
where  $\Delta= r^2 - \Lambda(u) r^4/3+a^2$, $R=r+ia\,{\rm cos}\,\theta$,  $R^2=r^2+a^2{\cos}^2\theta$ and $u$ is retarded time co-ordinate. This line element will represent rotating stationary de Sitter solution when $\Lambda(u)$ becomes constant \citep{ibo05}. Also if one sets the rotational parameter $a$ to zero, it will describe the non-stationary non-rotating de Sitter solution \citep{ibo09}. Here it is convenient to express the null tetrad vectors explicitly for further use as
\begin{eqnarray}
\ell_a&=&\delta^1_a - a{\ sin}^2\theta\ \delta^4_a,\cr
n_a&=&\frac{\Delta}{2\,R^2}\,\delta^1_a+ \delta^2_a
-\frac{\Delta}{2\,R^2}\,a\,{\rm sin}^2\theta\,\delta^4_a,\cr
m_a&=&-\frac{1}{\sqrt{2}R}\{-ia\,{\sin}\,\theta\ \delta^1_a
+R^2\delta^3_a \cr &&+i(r^2 +a^2){\sin}\,\theta\ \delta^4_a \}.
\end{eqnarray}
Here $\ell_a$,\, $n_a$ are real null vectors and $m_a$ is complex
with the normalization conditions $\ell_an^a= 1 = -m_a\bar{m}^a$ and other inner products vanish. The directional derivatives are defined by
\begin{eqnarray}
&&D=\partial_{r},\;\;\;\nabla=\frac{r^2+a^2}{R^2}\,\partial_{u}-\frac{\Delta}{2\,R^2}\,\partial_{r}
+\frac{a}{R^2}\,\partial_{\phi},\cr
&&\delta=\frac{1}{\sqrt{2}R}\{ia\,{\sin}\,\theta\
\partial_{u}+\partial_{\theta}+\frac{i}{{\sin}\,\theta}\,\partial_{\phi}\}.
\end{eqnarray}
Then, the NP spin-coefficients for the metric (1) are given as follows:
\begin{eqnarray}
&&\kappa=\sigma=\lambda=\epsilon=0,\cr
&&\rho=-\frac{1}{\bar{R}},\;\;\;\mu=-\frac{\Delta}{2\bar{R}R^2},\cr
&&\alpha=\pi-\bar{\beta},\;\;\;\beta=\frac{{\rm
cot}\,\theta}{2\sqrt{2}R},\cr&& \pi=\frac{ia\,{\
sin}\,\theta}{\sqrt{2}\bar{R}\bar{R}},\;\;\tau=-\frac{ia\,{\sin}\,\theta}{\sqrt{2}R^2}, \\
&&\gamma=\frac{1}{2\bar{R}R^2}\Big\{r\Big(1-\frac{2\Lambda(u)
r^2}{3}\Big)\bar{R}-\Delta\Big\},\cr
&&\nu=\frac{1}{6\sqrt{2}\bar{R}R^2}ia\,{\rm sin}\,\theta\ \Lambda(u),_u r^4. \nonumber
\end{eqnarray}
From these NP spin-coefficients, it is to mention that the rotating non-stationary
de Sitter model possesses null geodesic affinely parameterized $(\kappa=\epsilon=0)$,
shear free $(\sigma=0)$, expanding $(\hat{\theta}\neq0)$ and rotating
$(\hat{\omega}^{2}\neq0)$ null vector $\ell_a$.
\begin{eqnarray}
&&\hat{\theta}\equiv-\frac{1}{2}(\rho+\bar{\rho})=\frac{r}{R^2},\cr
&&\hat{\omega}^2\equiv-\frac{1}{4}(\rho-\bar\rho)^2=\frac{a^2{\cos}^2\theta}{R^2R^2}.
\end{eqnarray}

By the virtue of Einstein's field equations $R_{ab}-(1/2)R\,g_{ab}=-KT_{ab}$, the energy-momentum
tensor describing matter field for the space-time (1) is given as
\begin{eqnarray}
T_{ab} &=&\mu^*\,\ell_a\,\ell_b+ 2\,\rho^*\,\ell_{(a}\,n_{b)}
+2\,p\,m_{(a}\bar{m}_{b)}\cr&&+ 2\,\omega\,\ell_{(a}\,\bar
{m}_{b)} + 2\,\bar\omega\,\ell_{(a}\,m_{b)},
\end{eqnarray}
where the quantities are                                                                                                                                        \begin{eqnarray}
&&\mu^*=-\frac{r^4}{6KR^2R^2}\{2r\Lambda(u),_u +a^2{\sin}^2\theta\Lambda(u),_{uu}\},\cr
&&\rho^*=\frac{r^4}{KR^2R^2}\Lambda(u),\cr
&&p=-\frac{r^2}{KR^2R^2}\{r^2+2a^2{\cos}^2\theta\}\Lambda(u),\cr
&&\omega=-\frac{iar^3{\sin}\,\theta}{6\sqrt{2}KR^2R^2}(R-3\bar{R})\Lambda(u),_u,
\end{eqnarray}
with the universal constant $K=8\pi G/c^4$, \, $\mu^*$ describes null density arisen from the non-stationary property of the solution with cosmological function $\Lambda(u)$.  $\rho^*$ and $p$ are the
density and pressure of the matter distribution which does not involve
derivatives of cosmological function $\Lambda(u)$. Here,
$\omega$ represents the rotational density determined by the rotational parameter $a$ coupling with the derivative
of $\Lambda(u)$. However, $\omega$ will vanish when either $a$ is zero for {\it non-rotating}, or $\Lambda(u)$ becomes constant for {\it stationary} models.
The energy-momentum tensor  (6) satisfies the conservation equations, expressed in NP formalism \citep{ibo09}
\begin{equation}
T^{ab}_{\;\;\;; b}=0,
\end{equation}
which shows that the non-stationary de Sitter metric $(1)$ is an exact solution of Einstein's field equations.

We shall discuss the nature of time-like unit vector $u^a=(\sqrt{2})^{-1}(\ell_a+n^a)$. By knowing the nature of
time-like vector field of the observer, we can explain the physical properties of a matter distribution of rotating line element i.e., whether the field is expanding $(\Theta=u^a_{;a}\neq0)$, shearing $(\sigma_{ab}\neq0)$, rotating $(\omega_{ab}\neq0)$ or accelerating $(\dot{u}=u_{a;b}u^b)$. The expansion scalar $\Theta$ and acceleration vector $\dot{u}_{a}=u_{a;b}u^b$ are obtained as follows
\begin{eqnarray}
\Theta&=&\frac{r}{\sqrt{2}R^2}\Big\{1+\frac{2\Lambda(u) r^2}{3}\Big\},\cr
\dot{u}_a&=&\frac{r}{2R^2}\Big\{1-\frac{2\Lambda(u) r^2}{3}-\frac{\Delta}{R^2}\Big\}v_a\cr&&-\frac{a^2{\sin}^2\theta}{2\sqrt{2}R^2R^2}\Big(Rm_a+\bar{R}\bar{m}^a\Big).
\end{eqnarray}
where $v_a=\frac{1}{\sqrt{2}}(\ell_a-n_a)$ and $\Delta= r^2 - \Lambda(u)
r^4/3+a^2$. Then the shear $\sigma_{ab}$ and  vorticity $\omega_{ab}$ tensors are found as
\begin{eqnarray}
\sigma_{ab}&=&u_{(a;b)}-\dot{u}_{(a}u_{b)}-\frac{1}{3}\Theta (g_{ab}-u_au_b)\cr
&=&-\frac{1}{3\sqrt{2}R^2}\Big[\frac{3r\Delta}{R^2}-4(r-\frac{\Lambda(u) r^3}{3})\Big]\cr&&\times\Big(v_av_b-m_{(a}\bar{m}_{b)}\Big)
-\frac{1}{\sqrt{2}\bar{R}R^2}\cr&&\times\Big[2ai\,{\sin}\,\theta(2r+ia\,{\cos}\,\theta)-\cr&&\frac{i{\sin}\,\theta \Lambda(u)_{,u } r^4}{6}\Big]v_{(a}m_{b)}+\{c.c.\} \cr
\omega_{ab}&=&u_{[a;b]}-\dot{u}_{[a}u_{b]}\cr
&=&-\frac{ia\,{\sin}\,\theta}{2\sqrt{2}\bar{R}R^2}\Big[2ai\,{\cos}\,\theta+
\frac{\Lambda(u)_{,u} r^4}{6}\Big]v_{[a}m_{b]}\cr &&+\{c.c.\} +\frac{ai\,{\cos}\,\theta}{\sqrt{2}R^2}\Big(2+\frac{\Delta}{R^2}\Big)
m_{[a}\bar{m}_{b]},
\end{eqnarray}
which are orthogonal to $u^a$ i.e., $\sigma_{ab}u^b=0$ and $\omega_{ab}u^b=0$.
Here $\{c.c.\}$ denotes the complex conjugate of the previous term. The expansion scalar $\Theta$ and acceleration $\dot{u}_a$ in $(9)$ of the solution depend  on the function $\Lambda(u)$. The presence of rotational parameter $a$ in the rotating metric (1) shows directly the existence of  vorticity tensor in (10). When the rotational parameter $a$ sets to be zero, the vorticity tensor $\omega_{ab}$ will vanish, indicating the nature of the non-rotating solution.

The  non-stationary rotating de Sitter solution (1) has
singularity when
\begin{equation}
\Delta=r^2-{r^4\Lambda(u)}/3+a^2=0.
\end{equation}
This polynomial equation has four roots $r_{+\,+}$, $r_{+\,-}$, $r_{-\,+}$ and
$r_{-\,-}$. They are found as
\begin{eqnarray}
r_{\pm(\pm)}=\pm\sqrt{\frac{1}{2\Lambda(u)}\Big\{3\pm\sqrt{9
+12a^2\Lambda(u)}\,\Big\}}.
\end{eqnarray}
Now let us denote  these roots $r_{+\,+}$, $r_{+\,-}$, $r_{-\,+}$,
$r_{-\,-}$ as $r_1$, $r_2$, $r_3$, $r_4$ respectively for
simplicity, $r_j, \;\, j=1,2,3,4$. They satisfy the
following relation
\begin{eqnarray}
(r-r_1)(r-r_2)(r-r_3)(r-r_4)  \nonumber \\
= -\,\frac{3}{\Lambda(u)}\Big\{r^2-\frac{r^4\Lambda(u)}{3}+a^2\Big\}.
\end{eqnarray}
Then each root will describe the location of the cosmological horizons
for the observer and associates  an area of the horizon at each
point at $r=r_j, \;\, j=1,2,3,4$,
\begin{eqnarray*}
{\cal A}_j &=&
\int_{0}^{\pi}\int_{0}^{2\pi}\sqrt{g_{\theta\theta}g_{\phi\phi}}
\,d\theta\,d\phi\,\Big|_{r=r_{j}} \cr\cr &=& 4\pi\{r_{j}^2 +a^2\}.
\end{eqnarray*}

The horizons are necessarily null surfaces $r=r(u,\theta)$ that satisfy the null surface conditions
\begin{eqnarray*}
g^{ab}\partial_a F\partial_b F=0.
\end{eqnarray*}
The generalized tortoise co-ordinate transformation (GTCT) technique gives the location and temperature of horizons of non-stationary black hole \citep{wocai01}. Using GTCT, one can reduce Klein-Gordon or Chandrasekhar-Dirac equations to a standard form of wave equations near the horizon by introducing common tortoise co-ordinates transformation by $r_*=r+(2\kappa)^{-1}ln(r-r_H)$ in a non-stationary or stationary space-time, where $\kappa$ is the adjustable parameter and $r_H$ is the event horizon. We assume that the locations of the de Sitter horizons $r_j$ are functions of retarded time co-ordinates $u=t-r_*$ and angle $\theta$.

Here, since the rotating de Sitter cosmological space has the four roots, we would like to introduce an improved GTCT as follows:
 \begin{eqnarray}
&&r_*=r+\frac{1}{2\kappa_j(u_0, \theta_0)}ln[r-r_j(u,\theta)],\cr
&&u_*=u-u_0,\quad \theta_*=\theta-\theta_0,
 \end{eqnarray}
 where $r_j=r_j(u, \theta)$ are the locations of the horizons and  $\kappa_j$ are the adjustable parameters to be determined. All parameters $\kappa_j, u_0$ and $\theta_0$ are constant under the GTCT. From the above equation (14), we have the following operators
\begin{eqnarray}
&&\frac{\partial}{\partial r}=\frac{\partial}{\partial r_*}+\frac{1}{2\kappa_j(r-r_j)}\frac{\partial}{\partial r_*},\cr &&\frac{\partial}{\partial\theta}=\frac{\partial}{\partial\theta_*}-\frac{\acute{r}_j}{2\kappa_j(r-r_j)}\frac{\partial}{\partial r_*},\\
&&\frac{\partial}{\partial u}=\frac{\partial}{\partial u_*}-\frac{\dot{r}_j}{2\kappa_j(r-r_j)}\frac{\partial}{\partial r_*},\nonumber
\end{eqnarray}
where $\dot{r}_j=\partial r_j/\partial u$, $\acute{r}_j=\partial r_j/\partial\theta$. Now applying the GTCT in (14) to the null surface condition $g^{ab}\partial_aF\partial_bF=0$
and taking limit as $r\rightarrow r_j(u_0,\theta_0)$, $u\rightarrow u_0$ and $\theta\rightarrow\theta_0$, where $u_0$ and $\theta_0$ are the initial states of the cosmological horizons, then we get as
\begin{eqnarray}
\Big[\Delta_j +2(r^2_j+a^2)\dot{r}_j+a^2{\sin}^2\theta_0\dot{r}^2_j \nonumber \\+ \acute{r}^2_j\Big]\Big(\frac{\partial F}{\partial r_*}\Big)^2=0.
\end{eqnarray}
The above equation holds only when the content of the square bracket is equal to zero giving the location of horizons of rotating de Sitter cosmological model as
\begin{equation}
 \Delta_j +2(r^2_j+a^2)\dot{r}_j+a^2{\sin}^2\theta_0\dot{r}^2_j+\acute{r}^2_j=0,
\end{equation}
where $\Delta_j=r^2_j-\Lambda(u_0)r^4_j/3+a^2$. Here $\dot{r}_j=\partial r_j/\partial u$ is the rate of horizons with respect to time and $\acute{r}_j=\partial r_j/\partial\theta$ is the rate of change with angle. Equation $(17)$ shows that the shape of cosmological horizons depend on time and angle due to the GTCT transformation. Equation (17) has four roots as in (12) and they are given below:
\begin{eqnarray}
r_j&\equiv& r_{\pm(\pm)}\cr \cr &=&\pm\sqrt{\Big[{\frac{1}{2\Lambda(u)}\Big\{3 + 6\dot{r}_j  \pm\sqrt{W}\,\Big\}}}\Big],
\end{eqnarray}
where $W={9+12a^2\Lambda(u)+3H}$ and $H=12\dot{r}_j (1+\dot{r}_j) +4\Lambda(u)\{\acute{r}^2_j+2a^2\dot{r}_j
+a^2\dot{r}^2_j\sin^2\theta\}.$ These roots are due to the GTCT in (14) and will reduce to those of the non-stationary rotating de Sitter cosmological model given in (12) when $\acute{r}_j=\dot{r}_j=0$ in (18).

\section{Klein-Gordon equation}

 The wave equation describing the dynamical behavior of scalar particle of mass $\mu_0$ is given by
\begin{equation}
\frac{1}{\sqrt{-g}}(\partial_a\sqrt{-g}g^{ab}\partial_b\Phi)-\mu^2_0\Phi=0.
\end{equation}
For the de Sitter line element $(1)$, this equation reduce to
\begin{eqnarray}
&&\Big[\Delta \frac{\partial^2}{\partial r^2}-2(r^2+a^2)\frac{\partial^2}{\partial u\partial r}+2a\frac{\partial^2}{\partial u\partial\phi}\cr&&-2a\frac{\partial^2}{\partial\phi\partial r}+\frac{\partial^2}{\partial\theta^2}+\frac{1}{{\sin}^2\theta}\frac{\partial^2}{\partial \phi^2}+a^2{\sin}^2\theta\frac{\partial^2}{\partial u^2} \cr&&+\cot\theta\frac{\partial}{\partial\theta}-2r\frac{\partial}{\partial u}+\Big(2r-\frac{4\Lambda(u) r^3}{3}\Big)\frac{\partial}{\partial r}\cr&&+\mu^2_0R^2\Big]\Phi=0.
\end{eqnarray}
Under the GTCT and taking the limit $r\longrightarrow r_j(u_0,\theta_0)$, as $u \longrightarrow u_0$, $\theta\longrightarrow \theta_0$, the above equation can be written as
\begin{eqnarray}
&&\Big[\frac{1}{\kappa_j}\Big\{r_j-\frac{2\Lambda(u_0)r^3_j}{3}+2\dot{r}_j r_j\Big\}+2\Delta_j\cr&&+2(r^2_j+a^2)\Big]\frac{\partial^2\Phi}{\partial r^2_*}-2\acute{r}_j\frac{\partial^2\Phi}{\partial r_*\partial\theta_*}\cr&&-2a(1+\dot{r}_j)\frac{\partial^2\Phi}{\partial r_*\partial\phi}-2\Big\{(r^2_j+a^2)\cr&&+a^2{\sin}^2\theta_0\dot{r}_j\Big\}\frac{\partial^2\Phi}{\partial r_* \partial u_*}-(2r_j\dot{r}_j+a^2{\sin}^2\theta_0\ddot{r}_j\cr&&+\cot\theta_0\,\acute{r}_j+r^{''}_j)\frac{\partial\Phi}{\partial r_*}=0.
\end{eqnarray}
In deriving the above equation, we use L' Hospital rule to deal with infinite form of $0/0$ and  the value of the limit is given as
\begin{eqnarray}
 &&\lim_{r \rightarrow r_j}\frac{\Delta+2(r^2+a^2)\dot{r}_j+a^2{\sin}^2\theta_0\dot{r}^2_j+\acute{r}^2_j}{r-r_j}
\cr&&=2\Big[r_j-\frac{2\Lambda(u_0)r^3_j}{3}+2r_j\dot{r}_j\Big].
\end{eqnarray}
The standard form of wave equation near the horizons can be obtained by adjusting the parameter $\kappa_j$ for each $r_j$ in $(21)$. The wave equation (21) shall be discussed later in Section 6.

\section{Chandrasekhar-Dirac equations}

The four coupled Chandrasekhar-Dirac equations in Newman-Penrose formalism \citep{newpen62} are
given by
\begin{eqnarray}
&&(D+\epsilon-\rho)F_1 +
(\bar{\delta}+\pi-\alpha)F_2=i\mu_0G_1,\cr\cr
&&(\nabla+\mu-\gamma)F_2+(\delta+\beta-\tau)F_1=i\mu_0G_2, \cr\cr
&&(D+\bar{\epsilon}-\bar{\rho})G_2-(\delta+\bar{\pi}-\bar{\alpha})G_1=i\mu_0F_2,\cr\cr
&&(\nabla+\bar{\mu}-\bar{\gamma})G_1-(\bar{\delta}+\bar{\beta}-\bar{\tau})G_2=i\mu_0F_1,\nonumber \\
\end{eqnarray}
where $\mu_0$ denotes the mass of the Dirac particles and $F_1$, $F_2$, $G_1$ and $G_2$  are the four components of the wave function. By substituting the spin-coefficients from $(4)$ in $(23)$, we have the following equations
\begin{eqnarray}
(\partial_r+\frac{1}{\bar{R}})F_1+\frac{1}{\sqrt{2}\bar{R}}L^{-}F_2=i\mu_0G_1, \cr
-\frac{\Delta}{2R^2}D^{+}F_2+\frac{1}{\sqrt{2}R}(L^{+}+\frac{ia\,\sin\theta}{\bar{R}})F_1
=i\mu_0G_2,\cr
-\frac{1}{\sqrt{2}R}L^{+}G_1+(\partial_r+\frac{1}{R})G_2=i\mu_0F_2,\cr
-\frac{\Delta}{2R^2}D^{+}G_1-\frac{1}{\sqrt{2}\bar{R}}(L^{-}
+\frac{ia\sin\theta}{R})G_2=i\mu_0F_1,\nonumber \\
\end{eqnarray}
where operators are given by
\begin{eqnarray}
L^{+}&=&\partial_\theta+ia\,\sin\theta\ \partial_u
+\frac{i}{{\sin}\,\theta}\,\partial_\phi+\frac{\cot\theta}{2},\cr
L^{-}&=&\partial_\theta-ia\,\sin\theta\ \partial_u
-\frac{i}{\sin\theta}\,\partial_\phi+\frac{\cot\theta}{2},\cr
D^{+}&=&\partial_r-\Delta^{-1}\Big\{2(r^2+a^2)\,\partial_u+2a\,\partial_\phi\cr&&-r\Big(1-\frac{2\Lambda(u) r^2}{3}\Big)\Big\}.
\end{eqnarray}
We define $F_1$, $F_2$, $G_1$ and $G_2$ by
\begin{eqnarray}
&&F_1=\frac{f_1}{\sqrt{2}\bar{R}},\quad G_1=\frac{g_1}{\sqrt{2}},\cr
&&F_2=f_2,\quad G_2=\frac{g_2}{2R}.
\end{eqnarray}
Then equation $(24)$ becomes
\begin{eqnarray}
\frac{\partial f_1}{\partial r}+ L^{-}f_2&=&i\mu_0 \bar{R}g_1,\cr\cr
L^{+}f_1-\Delta D^{+}f_2 &=& i\mu_0 \bar{R}g_2,\cr\cr
-L^{+}g_1+\frac{\partial g_2}{\partial r}&=&2i\mu_0Rf_2,\cr\cr
\Delta D^{+}g_1 + L^{-}g_2&=&-2i\mu_0Rf_1.
\end{eqnarray}
 We may consider with a pair of components $f_1$, $f_2$ or $g_1$, $g_2$ only. In order to investigate the Hawking's radiation, we will consider equation $(27)$ near the horizons only. Under the transformation $(15)$ the limiting form of equation $(27)$ near the horizons are obtained as follows
\begin{eqnarray}
\frac{\partial f_1}{\partial r_{*}}-(\acute{r}_j-ia\dot{r}_j\,\sin\theta_0)\frac{\partial f_2}{\partial r_{*}}=0,\nonumber \\
\Big\{\acute{r}_{j}+\dot{r}_jia\,\sin\theta_0\Big\}\frac{\partial f_1}{\partial r_{*}}+\Big\{\Delta_{j}\nonumber \\+2\dot{r}_j(r^2_{j}+a^2)\Big\}\frac{\partial f_2}{\partial r_{*}}
=0,
\end{eqnarray}
and
\begin{eqnarray}
(\acute{r}_j+ia\dot{r}_j\,\sin\theta_0)\frac{\partial g_1}{\partial r_{*}}+\frac{\partial g_2}{\partial r_{*}}=0, \nonumber \\
\Big\{\Delta_{j}+2\dot{r}_j(r^2_{j}+a^2)\Big\}\frac{\partial g_1}{\partial r_{*}}
-\Big\{\acute{r}_{j}\nonumber \\-\dot{r}_jia\,\sin\theta_0\Big\}\frac{\partial g_2}{\partial r_{*}}=0,
\end{eqnarray}
after taking the limit as $r\rightarrow r_j(u_0,\theta_0)$, $u\rightarrow u_0$ and $\theta\rightarrow \theta_0$. If the four derivatives $\frac{\partial f_1}{\partial r_{*}}$, $\frac{\partial f_2}{\partial r_{*}}$, $\frac{\partial g_1}{\partial r_{*}}$ and
$\frac{\partial g_2}{\partial r_{*}}$ of the above equations (28), (29) are nonzero, then the condition for non-trivial solutions for  $f_1$, $f_2$, $g_1$ and $g_2$ is that the determinant of its coefficient will be zero. This gives the equation:
\begin{eqnarray*}
 \Delta_j +2(r^2_j+a^2)\dot{r}_j+a^2{\sin}^2\theta_0\dot{r}^2_j+\acute{r}^2_j=0,
\end{eqnarray*}
where $\Delta_j=r^2_j-\Lambda(u_0)r^4_j/3+a^2$. This equation is same as the equation for the null surface (17) given above. The equation $(28)$ can eliminate the crossing term of first order derivatives in second order Chandrasekhar-Dirac equations near $r=r_j$.

A direct calculation gives the second order Chandrasekhar-Dirac equations as follows
\begin{eqnarray*}
\Delta D^+\frac{\partial f_1}{\partial r}+L^-L^+f_1-2\mu^2_0R^2f_1
=i\mu_0\Delta g_1 \\-a\mu_0\sin\theta g_2  +\frac{ia\sin\theta r^3\dot{\Lambda}(u)}{3}\Big(r\frac{\partial f_2}{\partial r}+2f_2\Big),
\end{eqnarray*}
\begin{eqnarray}
\frac{\partial}{\partial r}(\Delta D^+f_2)+L^+L^-f_2-2\mu^2_0R^2f_2 \nonumber \\ =-a\mu_0\sin\theta g_1-i\mu_0 g_2.
\end{eqnarray}
Using the GTCT in $(30)$ and after some tedious calculations, the limiting form of above equation when $r$ tends to $r_j(u_0,\theta_0)$, $u$ approaches to $u_0$ and $\theta$ goes to $\theta_0$ leads to
\begin{eqnarray}
&&\Big\{2\Delta_j+2(r^2_j+a^2)\dot{r}_j+\frac{r_j+2r_j\dot{r}_j}{\kappa_j}\cr&&
-\frac{2\Lambda(u_0)r^3_j}{3\kappa_j}\Big\}\frac{\partial^2f_1}{\partial r^2_*}-2a(1+\dot{r}_j)\frac{\partial^2 f_1}{\partial r_*\partial\phi}\cr&&-2\acute{r}_j\frac{\partial^2 f_1}{\partial r_*\partial\theta_*}-2\Big\{(r^2_j+a^2)\cr &&
+a^2{\sin}^2\theta_0 \dot{r}_j\Big\}\frac{\partial^2f_1}{\partial r_*\partial u_*}-\Big\{r_j-\frac{2\Lambda(u_0) r^3_j}{3}\cr\cr &&+r^{''}_j +\acute{r}_j\cot\theta_0 -ia\cos\theta_0 \dot{r}_j \cr\cr &&+a^2{\sin}^2\theta_0\ddot{r}_j+4r_j\dot{r}_j -\frac{ia\sin\theta_0 \dot{\Lambda}(u_0)r^4_j}{3}\cr \cr &&\times\frac{\acute{r}_j+ia\sin\theta_0 \dot{r}_j}{\Delta_j+2\dot{r}_j(r^2_j+a^2)}\Big\}\frac{\partial f_1}{\partial r_*}=0 \nonumber \\
\end{eqnarray}
and
\begin{eqnarray}
&&\Big\{2\Delta_j+2(r^2_j+a^2)\dot{r}_j+\frac{r_j+2r_j\dot{r}_j}{\kappa_j}
\cr&&-\frac{2\Lambda(u_0)r^3_j}{3\kappa_j}\Big\}\frac{\partial^2f_2}{\partial r^2_*}-2a(1+\dot{r}_j)\frac{\partial^2 f_2}{\partial r_*\partial\phi}\cr&&-2\acute{r}_j\frac{\partial^2 f_2}{\partial r_*\partial\theta_*}\cr \cr &&-2\Big\{(r^2_j+a^2)+a^2{\sin}^2\theta_0 \dot{r}_j\Big\}\frac{\partial^2f_2}{\partial r_*\partial u_*}\cr\cr &&-\Big\{-r_j+\frac{2\Lambda(u_0) r^3_j}{3}+r^{''}_{j}+\acute{r}_j\cot\theta_0\cr&&-ia\cos\theta_0 \dot{r}_j +a^2{\sin}^2\theta_0\ddot{r}_j\Big\}\frac{\partial f_2}{\partial r_*}=0,
\end{eqnarray}
here the first-order derivative term $\partial_{r_*} f_2$ in equation $(31)$ is replaced by $\partial_{r_*} f_1$.
In order to reduce to a standard form of wave equation near the horizons, we adjust the parameters $\kappa_j$ such that they satisfy
\begin{eqnarray}
\frac{1}{\kappa_j}\Big\{r_j-\frac{2\Lambda(u_0) r^3_j}{3}+2r_j\dot{r}_j\Big\} +2(r^2_j+a^2)\dot{r}_j
\nonumber \\+2\Delta_j=r^2_j+a^2+a^2{\sin}^2\theta_0\dot{r}_j,
\end{eqnarray}
and using (17) we obtain the surface gravities of the horizons $r=r_j$ as
\begin{equation}
\kappa_j=\frac{r_j-\frac{2\Lambda(u_0) r^3_j}{3}+2r_j\dot{r}_j}{Q(1+2\dot{r}_j)+2\acute{r}^2_j},
\end{equation}
where $Q=r^2_j+a^2+a^2 {\sin}^2\theta_0\dot{r}_j$ with $j=1, 2, 3, 4$. Here we observe the involvement of the de Sitter cosmological function $\Lambda(u)$ at $u = u_0$ in the expression of surface gravities. These surface gravities $\kappa_j$ are the parameters for determining the temperature of the thermal radiation spectrum for scalar fields.

\section{Maxwell's equations}

In this section we shall analyze the Maxwell's equations in non-stationary rotating de Sitter space-time.
The four coupled Maxwell's equations for electromagnetic field in Newman-Penrose formalism \citep{newpen62} are given by
\begin{eqnarray}
D\phi_1-\bar{\delta}\phi_0&=&(\pi-2\alpha)\phi_0 +2\rho\phi_1\cr &&-\kappa\phi_2+2\pi_0 J_1,\cr
\delta\phi_2-\nabla\phi_1&=&-\nu \phi_0 +2\mu\phi_1 \cr &&+(\tau-2\beta)\phi_2+2\pi_0 J_2, \\
\delta \phi_1-\nabla \phi_0&=&(\mu-2\gamma)\phi_0+2\tau \phi_1\cr &&-\sigma \phi_2+2\pi_0 J_3,\cr
D\phi_2-\bar{\delta}\phi_1&=&-\lambda \phi_0+2\pi\phi_1\cr &&+(\rho-2\varepsilon)\phi_2+2\pi_0 J_4,\nonumber
\end{eqnarray}
where $\pi_0=(22/7)$ and $D=\ell^a\partial_a$, $\nabla=n^a\partial_a$, and $\delta=m^a\partial_a$ are the directional derivatives and $J_a$ is the tetrad components of the current vector given by $J_a=j_\mu Z^\mu_a$. In the above Maxwell's equations, by inserting the spin coefficients (4) and substituting $\phi_0=F_0$, $\phi_1=(\sqrt{2}\bar{R})^{-1}F_1$ and $\phi_2=2^{-1}\bar{R}^{-2}F_2$ to the above equation $(35)$,  we may write them as follows:
\begin{eqnarray}
&&\Big(\frac{\partial}{\partial r}+\frac{1}{\bar{R}}\Big)F_1-\Big(L_1-\frac{ia\sin\theta}{\bar{R}}\Big)F_0=0,\cr
&&\Big(\frac{\partial}{\partial r}-\frac{1}{\bar{R}}\Big)F_2-\Big(L_0+\frac{ai\sin\theta}{\bar{R}}\Big)F_1=0,\cr
&&\Big(L^+_0+\frac{ai\sin\theta}{\bar{R}}\Big)F_1+\Delta\Big(D^+_1-\frac{1}{\bar{R}}\Big)F_0=0,\cr
&&\Big(L^+_1-\frac{ai\sin\theta}{\bar{R}}\Big)F_2+\Delta\Big(D^+_0+\frac{1}{\bar{R}}\Big)F_1=0,\nonumber \\
\end{eqnarray}
where operators are defined by
\begin{eqnarray}
L_n&=&\partial_\theta-ia\sin\theta\ \partial_u
-\frac{i}{\sin\theta}\,\partial_\phi+n\cot\theta,\cr
L^+_n&=&\partial_\theta+ia\sin\theta\ \partial_u
+\frac{i}{\sin\theta}\,\partial_\phi+n\cot\theta,\cr
D^+_n&=&\partial_r-2\Delta^{-1}\Big\{(r^2+a^2)\partial_u+a\partial_\phi \cr&&-nr\Big(1-\frac{2\Lambda(u) r^2}{3}\Big)\Big\}.
\end{eqnarray}
Here $n$ is positive integer. The first order Maxwell's equations near the horizons are as follows
\begin{eqnarray}
\Big[\acute{r}_j-\dot{r}_j ia\sin\theta_0\Big]\frac{\partial F_0}{\partial r_*}+\frac{\partial F_1}{\partial r_*}=0,\nonumber \\
\Big\{\triangle_j+2(r^2_J+a^2)\dot{r}_j\Big\}\frac{\partial F_0}
{\partial r_*} -\Big\{\acute{r}_j \nonumber \\+ia\sin\theta_0\dot{r}_j\Big\}\frac{\partial F_1}{\partial r_*}=0,
\end{eqnarray}
and
\begin{eqnarray}
\Big[\acute{r}_j-\dot{r}_j ia\sin\theta_0\Big]\frac{\partial F_1}{\partial r_*}+\frac{\partial F_2}{\partial r_*}=0, \nonumber \\ \Big\{\triangle_j+2(r^2_j+a^2)\dot{r}_j\Big\}\frac{\partial F_1}{\partial r_*}  -\Big\{\acute{r}_j \nonumber \\ +ia\sin\theta_0\dot{r}_j\Big\}\frac{\partial F_2}{\partial r_*}=0.
\end{eqnarray}
We assume that the derivatives $\frac{\partial}{\partial r_*}F_0$, $\frac{\partial}{\partial r_*}F_1$ and $\frac{\partial}{\partial r_*}F_2$ are non-zero; and similar to the Dirac equations (28, 29), the condition for non-trivial solutions for  $F_0$, $F_1$ and $F_2$ are that the determinant of the coefficients of each (38), (39) will be zero. That provides the horizon equation like the Dirac equations or the null surface condition  (17).
Then, equation (39) can eliminate the crossing term in the second order Maxwell's equations near the horizons. The second order Maxwell's equations are given by
\begin{eqnarray*}
L^+_0L_1 F_0+\frac{\partial}{\partial r}(\Delta D^+_1F_0)-2R\frac{\partial F_0}{\partial u} =0,
\end{eqnarray*}
\begin{eqnarray*}
L^+_1L_0F_1+\frac{\partial}{\partial r}(\triangle D^+_0 F_1)+2R\frac{\partial F_1}{\partial u}+ \frac{2R F_1}{\bar{R}}
\nonumber \\+\frac{2a^2{\sin}^2\theta F_1}{\bar{R}\bar{R}}=\frac{4\Lambda(u)r^3 F_1}{3R}+\frac{2\Delta F_1}{\bar{R}\bar{R}},
\end{eqnarray*}
\begin{eqnarray}
\Delta D^+_0\frac{\partial F_2}{\partial r}+ L_0L^+_1F_2+ \frac{ia\sin\theta r^4 \dot{\Lambda}(u)}{3}\frac{\partial F_1}{\partial r}\nonumber \\+2R\frac{\partial F_2}{\partial u}=-\frac{ai\sin\theta \dot{\Lambda}(u)r^4 F_1}{3\bar{R}}.
\end{eqnarray}
The second order Maxwell's equations become
\begin{eqnarray}
&&\Big[\frac{1}{\kappa_j}\Big\{r_j-\frac{2\Lambda(u_0)r^3_j}{3}+2\dot{r}_j r_j\Big\}+2\Delta_j\cr&&+2\dot{r}_j(a^2+r^2_j)\Big]\frac{\partial^2F_0}{\partial r^2_*}
-2\acute{r}_j\frac{\partial^2F_0}{\partial r_*\partial\theta_*}\cr&&-2a(1+\dot{r}_j)\frac{\partial^2F_0}{\partial r_*\partial\phi}-2(r^2_j+a^2\cr&&+a^2{\sin}^2\theta_0\dot{r}_j)\frac{\partial^2F_0}{\partial u_*\partial r_*}+\Big[2r_j(1+\dot{r}_j)\cr&&-\frac{4\Lambda(u_0)r^3_j}{3}-r^{''}_j
-\acute{r}_j\cot\theta_0+2ai\cos\theta_0 \dot{r}_j\cr&&-a^2{\sin}^2\theta_0 \ddot{r}_j\Big]\frac{\partial F_0}{\partial r_*}=0,
\end{eqnarray}

\begin{eqnarray}
&&\Big[\frac{1}{\kappa_j}\Big\{r_j-\frac{2\Lambda(u_0)r^3_j}{3}+2\dot{r}_j r_j\Big\}+2\Delta_j\cr
&&+2\dot{r}_j(a^2+r^2_j)\Big]\frac{\partial^2F_1}{\partial r^2_*}-2\acute{r}_j\frac{\partial^2F_1}{\partial r_*\partial\theta_*}\cr
&&-2a(1+\dot{r}_j)\frac{\partial^2F_1}{\partial r_*\partial\phi}-2(r^2_j+a^2\cr&&+a^2{\sin}^2\theta_0\dot{r}_j)\frac{\partial^2F_1}{\partial \partial r_* \partial u_*}-\Big[a^2{\sin}^2\theta_0 \ddot{r}_j\cr&&+r^{''}_j+2r_j\dot{r}_j +\acute{r}_j\cot\theta_0\Big]\frac{\partial F_1}{\partial r_*}=0,
\end{eqnarray}

\begin{eqnarray}
&&\Big[\frac{1}{\kappa_j}\Big\{r_j-\frac{2\Lambda(u_0)r^3_j}{3}+2\dot{r}_j r_j\Big\}+2\Delta_j\cr
&&+2\dot{r}_j(a^2+r^2_j)\Big]\frac{\partial^2F_2}{\partial r^2_*}
-2\acute{r}_j\frac{\partial^2F_2}{\partial r_*\partial\theta_*} \cr &&-2a(1+\dot{r}_j)\frac{\partial^2F_2}{\partial r_*\partial\phi}-2(r^2_j+a^2  \cr &&+a^2{\sin}^2\theta_0\dot{r}_j)\frac{\partial^2F_2}{\partial r_* \partial u_*} -\Big[2r_j-\frac{4\Lambda(u_0) r^3_j}{3}\cr &&+4r_j\dot{r}_j
+r^{''}_j
+\acute{r}_j\cot\theta_0+a^2{\sin}^2\theta_0\ddot{r}_j
\cr &&+2(r_j+ia\cos\theta_0)\dot{r}_j
-\frac{ai\sin\theta_0 \dot{\Lambda}(u_0)r^4_j}{3}\cr &&\times\frac{\acute{r}_j+ia\sin\theta_0 \dot{r}_j}{\Delta_j+2(r^2_j+a^2)\dot{r}_j}\Big]\frac{\partial F_2}{\partial r_*}=0
\end{eqnarray}
after taking the limit as $r\rightarrow r_j(u_0,\theta_0)$, $u\rightarrow u_0$ and $\theta\rightarrow \theta_0$.
By reducing the second order Maxwell's equations under the transformation GTCT, the parameters $\kappa_j$ near the horizons are obtained as follows
\begin{equation}
\kappa_j=\frac{r_j-\frac{2\Lambda(u_0) r^3_j}{3}+2r_j\dot{r}_j}{Q(1+2\dot{r}_j)+2\acute{r}^2_j},
\end{equation}
which are the same as $(34)$ above -- the surface gravities obtained from Chandrasekhar-Dirac equations.

\section{Hawking temperature and thermal radiation spectra}

The Klein-Gordon equation $(21)$, Dirac equations ($31)$, $(32)$ and the second order Maxwell's equations $(41)$, $(42)$ and $(43)$ could be recast into a combined form near the horizons $r_j$ as follows:
\begin{eqnarray}
\frac{\partial^2\Psi}{\partial r^2_*}-2\Big\{K_1\frac{\partial^2\Psi}{\partial r_*\partial\theta_*}+\Omega_j\frac{\partial^2\Psi}{\partial r_*\partial\phi}+\frac{\partial^2\Psi}{\partial r_*\partial u_*}\nonumber \\-(K_2+iK_3)\frac{\partial\Psi}{\partial r_*}\Big\}=0,
\end{eqnarray}
where
\begin{eqnarray}
K_1=\frac{\acute{r}_j}{Q}, \quad
\Omega_j=\frac{1}{Q}a(1+\dot{r}_j),
\end{eqnarray}
where $Q=r^2_j+a^2+a^2 {\sin}^2\theta_0\,\dot{r}_j$.
The equation (45) is the standard form of wave equation in the non-stationary rotating de Sitter space. This wave equation includes the Klein-Gordon, Dirac and Maxwell's electromagnetic field equations having different conditions of the constant terms.

For example, when $\Psi=\Phi$ for Klein-Gordon scalar field, the equation (45) yields the following constants as
\begin{eqnarray}
K_2&=&-\frac{1}{2Q}(2r_j\dot{r}_j+a^2{\sin}^2\theta_0\ddot{r}_j\cr &&+\cot\theta_0 \acute{r}_j+r^{''}_j),\cr
K_3&=&0.
\end{eqnarray}
For Dirac particles when $\Psi=f_1$, the constants are given by
\begin{eqnarray}
K_2&=&\frac{1}{2Q}\Big[-r^{''}_j-\cot\theta_0 \acute{r}_j-r_j\cr &&+\frac{2\Lambda(u_0) r^3_j}{3}-a^2{\sin}^2\theta_0\ddot{r}_j-4r_j\dot{r}_j\cr
&&-\frac{\dot{r}_jr^4_j a^2{\sin}^2\theta_0 \dot{\Lambda}(u_0)}{3\{\Delta_j+2\dot{r}_j(r^2_j+a^2)\}}\Big],\cr
K_3&=&\frac{a}{2Q}\Big[ \cos\theta\dot{r}_j + \frac{\acute{r}_jr^4_j {\sin}\theta_0 \dot{\Lambda}(u_0)}{3\{\Delta_j+2\dot{r}_j(r^2_j+a^2)\}}\Big],\nonumber\\
\end{eqnarray}
and for $\Psi=f_2$, we get
\begin{eqnarray}
K_2&=&\frac{1}{2Q}\Big[r_j-\frac{2\Lambda(u_0) r^3_j}{3}-r^{''}_j\cr&&-\acute{r}_j\cot\theta_0-a^2\ddot{r}_j{\sin}^2\theta_0\Big],\cr
K_3&=&\frac{1}{2Q}a\cos\theta_0 \acute{r}_j.
\end{eqnarray}
Similarly, for Maxwell's electromagnetic field $\Psi=F_0$, the equation (45) provides the following constants:
\begin{eqnarray}
K_2&=&\frac{1}{Q}\Big[r_j(1+\dot{r}_j)-\frac{2\Lambda(u_0)r^3_j}{3}\cr&&-\frac{r^{''}_j}{2}
-\frac{\acute{r}_j\cot\theta_0}{2}-\frac{a^2{\sin}^2\theta_0 \ddot{r}_j}{2}\Big],\cr
K_3&=&\frac{a\cos\theta_0 \dot{r}_j}{Q};
\end{eqnarray}
for $\Psi=F_1$,
\begin{eqnarray}
K_2&=&-\frac{1}{2Q}(a^2{\sin}^2\theta_0\ddot{r}_j+r^{''}_j \cr &&+2r_j\dot{r}_j+
\acute{r}_j\cot\theta_0),\cr
K_3&=&0;
\end{eqnarray}
and for $\Psi=F_2$,
\begin{eqnarray}
K_2&=&-\frac{1}{Q}\Big[r_j-\frac{2\Lambda(u_0)r^3_j}{3}+3r_j\dot{r}_j
+\frac{r^{''}_j}{2}\cr &&+\frac{\acute{r}_j\cot\theta_0}{2}
+\frac{a^2{\sin}^2\theta_0 \ddot{r}_j}{2}\cr &&+\frac{a^2{\sin}^2\theta_0 \dot{r}_j r^4_j \dot{\Lambda}(u_0)}{6\{\triangle_j+2(r^2_j+a^2)\dot{r}_j\}}\Big],\cr
K_3&=&-\frac{a}{Q}\Big[\cos\theta_0\dot{r}_j-\frac{\acute{r}_j\sin\theta_0 \dot{\Lambda}(u_0)r^4_j}{6\{\triangle_j+2(r^2_j+a^2)\dot{r}_j\}}\Big],\nonumber \\
\end{eqnarray}
where $K_1$, $\Omega_j$, $K_2$ and $K_3$ are considered as real constants under the generalized coordinate transformation (14). One can view that equation $(45)$ is a second order partial differential equation representing wave equations near the horizons since all the coefficients $K_1$, $\Omega_j$, $K_2$ and $K_3$  are taken as real constants when $r\rightarrow r_j(u_0,\theta_0)$, $u\rightarrow u_0$ and $\theta\rightarrow \theta_0$.

Now following \citet{wocai01}, we may separate the variables in (45) for the analysis of the field equations as
\begin{equation}
\Psi=\hat{R}(r^*)\Theta(\theta^*){\rm e}^{i(m\phi-\tilde{\omega} u_*)}.
\end{equation}
The radial and angular parts of equation $(45)$ are respectively given by
\begin{eqnarray}
&&\Theta^{'}=\lambda {\Theta} \nonumber \\
&&\frac{\partial^2\hat{R}}{\partial r^2_*}-2\{im\Omega_j-i\tilde{\omega}-iK_3-K_0\}\frac{\partial\hat{R} }{\partial r_*}=0, \nonumber \\
\end{eqnarray}
where $\lambda$ is a real constant of the variable separation, $K_0=K_2-K_1\lambda$. The ingoing and outgoing wave solutions to $(45)$ near the horizons are given by
\begin{eqnarray}
\Psi^{\rm in}_\phi&\sim&e^{-i\tilde{\omega} u_*+im\phi+\lambda\theta_*},\cr
\Psi^{\rm out}_{\tilde{\omega}}&\sim& e^{-i\tilde{\omega} u_*+im\phi+\lambda\theta_*}(r-r_j)^{\frac{i(m\Omega_j
-\tilde{\omega}-K_3)}{\kappa_j}}
\cr&&\times(r-r_j)^{-\frac{K_0}{\kappa_j}}, \quad r>r_j,
 \end{eqnarray}
where $\Psi^{\rm in}_{\tilde{\omega}}$ represents an incoming wave and is analytic on $r=r_j$;  $\Psi^{\rm out}_{\tilde{\omega}}$ denotes an outgoing wave having logarithmic singularity on the horizons, but not analytic at $r=r_j$. Following \citet{dam76} and \citet{san88}, it can analytically continue from the outside of the horizon $r_j$ into its inside as
\begin{equation}
(r-r_j)\rightarrow \Big|r-r_j\Big|e^{-i\pi}=(r_j-r)e^{-i\pi}
\end{equation}
to
\begin{eqnarray}
\tilde{\Psi}^{\rm out}_{\tilde{\omega}} &\sim &e^{-i{\tilde{\omega}} u_*+im\phi+\lambda\theta_*}e^{\frac{\pi}{\kappa_j}(m\Omega_j-\tilde{\omega}-K_3)}\cr&& \times (r_j-r)^{\frac{-K_0+i(m\Omega_j-\tilde{\omega}-K_3)}{\kappa_j}}
e^{\frac{i\pi K_o}{\kappa_j}},
\end{eqnarray}
when $(r<r_j)$.
The relative scattering probabilities at $r=r_{j}$ are given by
\begin{equation}
\Big|\frac{\Psi^{\rm out}_{\tilde{\omega}} (r>r_j)}{\tilde{\Psi}^{\rm out}_{\tilde{\omega}}
(r<r_j)}\Big|^2=e^{\frac{-2\pi}{\kappa_j}(m\Omega_j-\tilde{\omega}-K_3)}.
\end{equation}
where $m$ is the azimuthal quantum number, $\tilde{\omega}$ is the energy of the radiating particles and $\Omega_j$ is the angular velocities near $r=r_{j}$. According to  the suggestion of \citet{dam76} and extended by \citet{san88}, the thermal radiation spectrum is given by
\begin{equation}
\langle N_{\tilde{\omega}} \rangle =\Big\{{\rm exp}\Big(\frac{m\Omega_j-\tilde{\omega}-K_3}{T_j}\Big)+1\Big\}^{-1},
\end{equation}
with
\begin{equation}
T_j=\frac{1}{2\pi}\Big\{\frac{r_j-\frac{2\Lambda(u_0)
r^3_j}{3}+2r_j\dot{r}_j}{Q(1+2\dot{r}_j)+2\acute{r}^2_j}\Big\},
\end{equation}
which is the Hawking's temperature defined by $T_j=\kappa_j/2\pi$. This determines the temperature of the thermal radiation near the horizons $r_j$ due to Klein-Gordon scalar field, Dirac particles and Maxwell's electromagnetic field  in the non-stationary rotating de Sitter cosmological space-time. Since the surface gravities $\kappa_j$ are same for these field equations, the temperatures of thermal radiation spectra are same as in (60) depending on the de Sitter cosmological function $\Lambda(u_0)$ near the horizons. However, it is observed that the thermal radiation spectra are different for different wave equations as the constant $K_3$ are not the same for all the scalar fields discussed here.

\section{Description of the
motion of rotating de Sitter metric in two dimensional space}

The line element of rotating de Sitter solution in Boyer-Lindquist
co-ordinate can be written as
\begin{eqnarray}
ds^2&=&\frac{\Delta-a^2{\sin}^2\theta}{R^2}dt^2-\frac{R^2}{\Delta}dr^2-R^2d\theta^2
\cr&&+\frac{2a{\sin}^2\theta}{R^2}\{(r^2+a^2)-\Delta\}dt d\phi +\frac{{\sin}^2\theta}{R^2}.\cr\cr &&\{a^2{\sin}^2\theta\Delta-(r^2+a^2)^2\}d\phi^2,
\end{eqnarray}
where $\Delta=r^2-\Lambda(u) r^4/3+a^2$ and
$R^2=r^2+a^2{\cos}^2\theta$. In order to describe the space-time geometry of the rotating de Sitter metric in two dimensional
space, we take $\theta=0, \pi$ and $d\theta=d\phi=0$ in the
equation $(61)$ and the two dimensional line element can be written
as
\begin{equation}
 ds^2=\Big(\frac{\Delta}{R^2}\Big)dt^2 -\Big(\frac{R^2}{\Delta}\Big)dr^2,
\end{equation}
where $R^2=r^2 + a^2$ and $f=\Delta R^{-2}$. On the symmetric axis, horizon is defined by
$g_{tt}=0$, whereas in general it defines the ergosurface. The two
dimensional line element could not describe the angular momentum and the transverse degree of freedom are simply redshifted away relatively to the ones in the $rt$-plane. To calculate the stress energy tensor for the metric $(62)$,
we transform to the double null form as
\begin{equation}
 ds^2=\Big(\frac{\Delta}{R^2}\Big)du dv,
\end{equation}
where $u=t-r^*, v=t+r^*$, where $r^*$ is defined as tortoise
coordinate given by $dr^*/dr=R^2/\Delta$. We obtained stress
energy components \citep{birdev86} for two dimensional non-stationary rotating de Sitter space as follows
\begin{eqnarray}
T_{uu}&=&T_{vv}=-\frac{1}{192\pi}\Big[\frac{\Delta,^2_r-4\Delta\{1-2\Lambda(u)
r^2\}}{R^2R^2}\cr\cr &&+\frac{4(\Delta \Delta,_r r
+\Delta^2)}{R^2R^2R^2}-\frac{12\Delta^2 r^2}{R^2R^2R^2R^2}\Big],\cr
T_{uv}&=&\frac{1}{48\pi}\Big[\frac{\Delta\{1-\Lambda(u)
r^2\}}{R^2R^2}-\frac{\Delta(2r\Delta,_r
+\Delta)}{R^2R^2R^2}\cr\cr&&+\frac{4r^2\Delta^2}{R^2R^2R^2R^2}\Big],\cr
T_{vv}&=&-\frac{1}{192\pi}\Big[\frac{\Delta,^2_r-4\Delta\{1-2\Lambda(u)
r^2\}}{R^2R^2}\cr\cr &&+\frac{4(\Delta \Delta,_r r
+\Delta^2)}{R^2R^2R^2}-\frac{12\Delta^2 r^2}{R^2R^2R^2R^2}\Big],\cr
T_{tt}&=&-\frac{1}{96\pi}\Big[\frac{\Delta,^2_r-8\Delta(1-2\Lambda(u)
r^2)}{R^2R^2}\cr\cr &&+\frac{8\Delta(2r\Delta,_r
+\Delta)}{R^2R^2R^2}-\frac{28r^2\Delta^2}{R^2R^2R^2R^2}\Big],\cr
T_{tr}&=&T_{rt}=0.
\end{eqnarray}
From above, we observed that there is no constant terms involved
in the expression $T_{uu}$. The Hawking flux can be obtained by considering the
value of $T_{uu}$ at $r_{++}$, the energy flux through the cosmological
horizon, where $r_{++}= \sqrt{\frac{3+\sqrt{9-
 12a^2\Lambda(u)}}{2\Lambda}}$. Since $\Delta=0$, $T_{vv}$ is given
 by
\begin{equation}
T_{vv}|_{r=r_{++}}=-\frac{1}{48\pi}\frac{r^2_{++}\Big(1-\frac{2\Lambda(u)
r^2_{++}}{3}\Big)^2}{(r^2_{++}+a^2)^2},
\end{equation}
which is exactly the negative of the Hawking flux. The temperature defined by $T_j=\kappa_j/2\pi$
of the non-stationary rotating de Sitter solution at the
cosmological horizons is given by
\begin{equation}
T=-\frac{\Lambda(u) r^3_{++}}{6\pi}\Big\{\frac{(r^2_{++}+2a^2)}{(r^2_{++}+a^2)^2}\Big\}.
\end{equation}
In the  above equation, the negative sign denotes the radiation of the energy-momentum tensor flux is radiated at infinity.

\section{Conclusions}
In this paper we have studied the Hawking's thermal radiation effects of Klein-Gordon, Dirac and Maxwell's electromagnetic fields in the non-stationary rotating de Sitter background space under the GTCT transformation. It is interesting to emphasis some of the physical properties of the de Sitter solution (1). The non-stationary rotating de Sitter model possesses the Weyl scalars  $\psi_2=\psi_3=\psi_4\neq0$ \citep{ibo09}. These non-zero Weyl scalars determine the gravitational field of the space-time that the de Sitter solution (1) is an algebraically special type II [$C_{abc[d}\ell_{h]}\ell^{b}\ell^{c}=0$, with $\psi_0=\psi_1=0$
\citep{chan83}] in the Petrov classification of space-time with a null vector
$\ell_a$.  The null vector $\ell_a$ is geodesic, shear free, expanding
($\hat{\theta}\neq 0$) as well as
non-zero twist ($\hat{\omega}\neq 0$) given in (5). Here the function $\Lambda(u)$,
describing the non-stationary status of the solution, does not involve in the expressions of expansion $\hat{\theta}$ and twist $\hat{\omega}$ in (5). This means
that the physical properties of this null vector $\ell^a$ are same
for both {\sl stationary} (with constant $\Lambda$) as well as {\sl non-stationary} (with function $\Lambda(u)$)
rotating de Sitter models, though they have different
gravitational fields possessing different energy momentum tensors.

The equations (17) and (34) above determine the locations and surface gravities of  non-stationary rotating de Sitter cosmological model, and their values depend on both the retarded time $u$ and angle $\theta$.  Under the  GTCT transformation, the second order equations of the Klein-Gordon, Dirac and Maxwell's electromagnetic fields are transformed to the standard wave equation near $r = r_j$, and separation of variables can be done as in (53). It is observed that the constant $K_3$ is absent in the spectrum of Klein-Gordon scalar particles in (47). Also $K_3$ of Dirac and Maxwell's equations will be vanished when the non-stationary de Sitter solution (1) becomes stationary with $\dot{r}_j=\acute{r}_j=0$ or it has a zero angular velocity with $a=0$. We have also seen the effect of the cosmological function $\Lambda(u)$ in the temperatures of thermal radiation spectra for all the three wave functions under the GTCT transformation in non-stationary de Sitter background. For a non-stationary non-rotating line element with $a = 0$, the constants $\Omega_j$, $K_3$ in both wave equations become zero, and the other constants will not vanish. For stationary non-rotating metric, the constant $K_3$ will be zero for wave equations. The surface gravities for Klein-Gordon scalar field, Dirac particles and Maxwell's electromagnetic field equations under the transformation of GTCT are the same.

It is interesting to mention that the angular velocities $\Omega_j$ given in (46) are due to the GTCT transformation in the non-stationary rotating de Sitter cosmological space:
\begin{equation}
\Omega_j=\frac{a(1+\dot{r}_j)}{r^2_j+a^2+a^2 {\sin}^2\theta_0\,\dot{r}_j}.
\end{equation}
The appearance of $\dot{r}_j$ in this equation indicates the effect of GTCT in the rotating de Sitter space. The rotating de Sitter cosmological space-time, stationary or non-stationary, will have the same angular velocities $\Omega_j$ under the GTCT transformation. When $\dot{r}_j=0$, $\Omega_j$ will be those of rotating de Sitter space before the transformation as
\begin{equation}
\Omega=\frac{a}{r^2+a^2}
\end{equation}
which depends on the roots $r=r_j$, $j=1,2,3,4$ of the polynomial $\Delta=0$. It is observed that the angular velocity for every rotating space-time, stationary or non-stationary, has the same structure as in (68). But it depends on the roots of the polynomial.

For example, the rotating Kerr black hole has the angular velocity $\Omega=a(r^2+a^2)^{-1}$, where $r=M\pm\surd{(M^2-a^2)}$; similarly  Kerr-Newman having $r=M\pm\surd{(M^2-a^2-e^2)}$ with the charge $e$; Kerr-Newman-Vaidya \citep{ibokap10} having $r=M+f(u)\pm\surd{\{(M+f(u))^2-a^2-e^2\}}$, where $f(u)$ is the Vaidya mass; rotating Vaidya-Bonnor, $r=f(u)\pm\surd{\{f^2(u)-a^2-e^2(u)\}}$ with the charge $e(u)$ and so on. From the study of Hawking's radiation effect of Klein-Gordon scalar field, Dirac particles and Maxwell's electromagnetic field in the non-stationary rotating de Sitter cosmological space-time under the GTCT, it is also found that the equation (45) could be considered as a general standard wave equation to be tackle for every non-stationary rotating space. The example of this situation is that three different wave equations, namely Klein-Gordon, Dirac and Maxwell could be recast into this standard form (45) with the variation in the values of the constants $K_2, K_3$. Also all these three wave equations have the same thermal radiation temperature (60) with the parameters $\kappa_j$ (33) or (44) for the non-stationary rotating de Sitter cosmological space-time.

The difficulty in the separation of variables of the standard wave equation (45) may be seen that if one assumes
\begin{equation}
\Psi=\hat{R}(r^*)\Theta(\theta^*)\hat{\Psi}(\phi){\rm exp}(-i\tilde{\omega} u_*),
\end{equation}
in (53), then the relative scattering probabilities at $r=r_{j}$ are given by
\begin{equation}
\Big|\frac{\Psi^{\rm out}_{\tilde{\omega}} (r>r_j)}{\tilde{\Psi}^{\rm out}_{\tilde{\omega}}
(r<r_j)}\Big|^2=e^{\frac{-2\pi}{\kappa_j}(\tilde{\beta}-\tilde{\omega}-K_3)},
\end{equation}
where $\tilde{\beta}$  is the separation constant $Z=\tilde{\alpha}+i\tilde{\beta}$. The equation (70) does not involve the angular velocities $\Omega_j$ given in (46) or (67), which shows the difference from the scattering probabilities (58) with the separation of variables (53). The equations (53) and (69) show at least two possibilities for the separation of variables in non-stationary space-times. However, it seems that the separation in (53) may be a convenient one, because it shows the direct involvement of the angular velocities $\Omega_j$ without disturbing the rotational character of the background space-time.

The investigation of Hawking's radiation effect of Klein-Gordon scalar field, Dirac particles and Maxwell's electromagnetic field in non-stationary rotating de Sitter space with cosmological function $\Lambda(u)$ is not been seen discussed before. The work discussed here in the non-stationary rotating de Sitter model includes the results of the following models: (i) stationary rotating de Sitter when $\Lambda(u)$ becomes constant \citep{ibo05}, (ii) non-stationary non-rotating de Sitter when $a=0$ \citep{ibo09} and (iii) stationary non-rotating when $\Lambda(u)$ = constant and $a=0$ \citep{hawell73}.

{\bf Acknowledgement}: The work of Ibohal was supported by University Grant Commission (UGC), New Delhi, File No. 31-87/2005 (SR), and Ibungochouba acknowledges the CSIR for providing financial assistance as Junior Research Fellowship.

\end{document}